\def\papertitle{Optimization Techniques for a Physical Model of Human Vocalisation}
\def\paperauthorA{Mateo Cámara}
\def\paperauthorB{Zhiyuan Xu}
\def\paperauthorC{Yisu Zong}
\def\paperauthorD{José Luis Blanco}
\def\paperauthorE{Author Five}

\documentclass[twoside,a4paper]{article}
\usepackage{etoolbox}
\usepackage{dafx_23}
\usepackage{multirow}
\usepackage[table]{xcolor}
\usepackage{amsmath,amssymb,amsfonts,amsthm}
\usepackage{euscript}
\usepackage[T1]{fontenc}
\usepackage[utf8]{inputenc}
\usepackage{ifpdf}
\usepackage[english]{babel}
\usepackage{caption}
\usepackage{subfig} 
\usepackage{color}

\definecolor{darkgreen}{HTML}{13Ba07}

\input glyphtounicode
\ninept

\newcounter{numauth}
\newcounter{listcnt}
\newcommand\authcnt[1]{\ifdefined#1 \stepcounter{numauth} \fi}

\newcommand\addauth[1]{
\ifdefined#1 
\stepcounter{listcnt}
\ifnum \value{listcnt}<\value{numauth}
\appto\authorslist{, #1}
\else
\appto\authorslist{~and~#1}
\fi
\fi}
\authcnt{\paperauthorB}
\authcnt{\paperauthorC}
\authcnt{\paperauthorD}
\authcnt{\paperauthorE}
\authcnt{\paperauthorF}
\authcnt{\paperauthorG}
\authcnt{\paperauthorH}
\authcnt{\paperauthorI}
\authcnt{\paperauthorJ}
\def\authorslist{\paperauthorA}
\addauth{\paperauthorB}
\addauth{\paperauthorC}
\addauth{\paperauthorD}
\addauth{\paperauthorE}
\addauth{\paperauthorF}
\addauth{\paperauthorG}
\addauth{\paperauthorH}
\addauth{\paperauthorI}
\addauth{\paperauthorJ}

\usepackage{times}

\newif\ifpdf
\ifx\pdfoutput\relax
\else
   \ifcase\pdfoutput
      \pdffalse
   \else
      \pdftrue
\fi

\ifpdf 
  \usepackage[pdftex,
    pdftitle={\papertitle},
    pdfauthor={\authorslist},
    pdfsubject={Proceedings of the 26th International Conference on Digital Audio Effects (DAFx23)},
    colorlinks=false, 
    bookmarksnumbered, 
    pdfstartview=XYZ 
  ]{hyperref}
  \usepackage[pdftex]{graphicx}
\else 
  \usepackage[dvips]{epsfig,graphicx}
  \usepackage[dvips,
    pdftitle={\papertitle},
    pdfauthor={\authorslist},
    pdfsubject={Proceedings of the 26th International Conference on Digital Audio Effects (DAFx23)},
    colorlinks=false, 
    bookmarksnumbered, 
    pdfstartview=XYZ 
  ]{hyperref}
\fi
\usepackage[hypcap=true]{caption}
\title{\papertitle}

\fiveaffiliations{
\paperauthorA \,\thanks{\vspace{-3mm}}}
{\href{https://iptc.upm.es/}{Information Processing \& Telecomm. Center} \\ Universidad Politécnica de Madrid\\ Madrid, Spain\\
{\tt \href{mailto:mateo.camara@upm.es}{mateo.camara@upm.es}}\vspace{.5cm}
}
{\paperauthorB \,}
{\href{https://c4dm.eecs.qmul.ac.uk/}{Centre for Digital Music} \\ Queen Mary University of London \\ London, UK \\ {\tt \href{mailto:zhiyuan.xu@qmul.ac.uk}{zhiyuan.xu@qmul.ac.uk}}
}
{\paperauthorC \,}
{\href{https://c4dm.eecs.qmul.ac.uk/}{Centre for Digital Music} \\ Queen Mary University of London \\ London, UK \\ {\tt \href{mailto:y.zong@qmul.ac.uk}{y.zong@qmul.ac.uk}}
}
{\paperauthorD \,}
{\href{https://iptc.upm.es/}{Information Processing \& Telecomm. Center} \\ Universidad Politécnica de Madrid\\ Madrid, Spain\\ {\tt \href{mailto:jl.blanco@upm.es}{jl.blanco@upm.es}}
}
{\href{https://c4dm.eecs.qmul.ac.uk/}{Centre for Digital Music} \\ Queen Mary University of London \\ London, UK \\ {\tt \href{mailto:y.zong@qmul.ac.uk}{joshua.reiss@qmul.ac.uk}}
}

\begin{document}
\ifpdf 
  \DeclareGraphicsExtensions{.png,.jpg,.pdf}
\else  
  \DeclareGraphicsExtensions{.eps}
\fi


\maketitle

\begin{abstract}
We present a non-supervised approach to optimize and evaluate the synthesis of non-speech audio effects from a speech production model. We use the Pink Trombone synthesizer as a case study of a simplified production model of the vocal tract to target non-speech human audio signals --yawnings. We selected and optimized the control parameters of the synthesizer to minimize the difference between real and generated audio. We validated the most common optimization techniques reported in the literature \textcolor{black}{and a specifically designed neural network. We} evaluated several popular quality metrics as error functions. These include both objective quality metrics and subjective-equivalent metrics. We compared the results in terms of total error and computational demand. Results show that genetic and swarm optimizers outperform least squares algorithms at the cost of executing slower and that specific combinations of optimizers and audio representations offer significantly different results. The proposed methodology could be used in benchmarking other physical models and audio types.
\end{abstract}

\section{Introduction}
\label{sec:intro}

Articulatory synthesis provides a unique opportunity to delve into the mechanics of speech production \cite{kello2004neural, aryal2016data}. Unlike black box models, physical models achieve an \textit{interpretable representation} of the inner characteristics of the vocal tract. This allows for a deeper understanding of the processes involved in speech production. They also provide precise control of the speech's articulatory, resonance, and phonatory characteristics, such as the position of the tongue, lips, existing constrictions, or nose size; as well as informed control of model parameters. This makes natural-sounding synthetic speech \textcolor{black}{samples} less prone to artifacts than other synthesis models. Furthermore, they may produce any type of human sound coming out of the mouth and the nose. Those include sounds that are not words, such as sighs, laughs, yawns, and so on.

These \textit{non-speech sounds} are becoming increasingly important in today's audiovisual productions and digital interactions. From the sound effects in movies and videogames to the soundscapes in podcasts and audiobooks \cite{afsar2021generating, hsu2022synthesizing, anikin2019soundgen}, the ability to generate these sounds has become a critical aspect to produce realistic performances. Analyzing the ability of models to construct these types of sounds is crucial to understand the limits and limitations of models \cite{oord2016wavenet}, as well as the underlying complexities of producing naturally sounding \textcolor{black}{audio samples}. Answering those questions opens up new possibilities for sound and user-experience designers, video-game developers, and audio production professionals looking for new and innovative ways to create high-quality, realistic, and engaging sound experiences.

Physical models for speech synthesis pose challenges that are extensively reported in the literature. They often include many parameters that are difficult to configure simultaneously to achieve high-quality sounds. Their combined optimization can be demanding, computationally expensive, time-consuming, and challenging to implement in real time. These complications explain the need to improve and optimize the synthesizer.

Our research focuses on articulatory parameters from a black-box point of view. We optimize synthesizers without paying specific attention to what each parameter represents to maximize \textcolor{black}{objective} similarity by minimizing the difference between a target signal and the synthesized signal. This ensures superior generalization capabilities for the proposed method and valuable results for other contexts.

In this contribution, we look at the physical model known as the \textit{Pink Trombone (PT)}\footnote{https://dood.al/pinktrombone/}. This is a simplified version of the vocal tract that uses a small set of fundamental parameters to control the shape and movements of the articulators during speech production \cite{story2005parametric}. We fixed its articulatory bounds to focus on sounds that a human could physically produce, and used these to optimize the PT and compare its result with human \textcolor{black}{audio samples}. 

We conduct a case study using synthetic, sustained, and yawning sounds to understand its capabilities and limitations. We test different black-box strategies to predict the synthesizer control parameters, including well-known \textit{optimization techniques} and \textit{Deep Neural Networks}, trained on a set of PT synthetic \textcolor{black}{samples}. For experimentation, we use sound files generated by the PT, as well as \textcolor{black}{audio clips} downloaded from the Freesound platform \cite{2797}.

Experiments shall lay the foundations for studies on articulatory and production models with multiple parameters and different audio types. The dataset, test sounds, and algorithms are available online\footnote{https://slash-trombone.github.io/}. Our goals for this contribution are the following:

\begin{itemize}
    \item \textit{Determine if PT parameters can be accurately predicted exclusively from acoustic features}. We optimize synthesizer control variables from audio \textcolor{black}{samples} as a black-box.
    \item \textit{Determine best optimization technique for articulatory variables}. We evaluate how different optimizers perform in front of increasingly challenging sounds.
    \item \textit{Determine the error metric that yields a more satisfactory outcome.} We benchmark different techniques for standard error metrics and acoustic parameterizations of audio files.

\end{itemize}

The remainder of this paper is organized as follows. Sec. 2 expands on the optimization techniques and the parameterizations reported in the literature. Sec. 3 describes the experiments covered to meet our objectives. Sec. 4 analyzes and discusses the results obtained, and Sec. 5 concludes the paper.

\section{Background}
\begin{figure}[t]
\centering
\includegraphics[width=68mm]{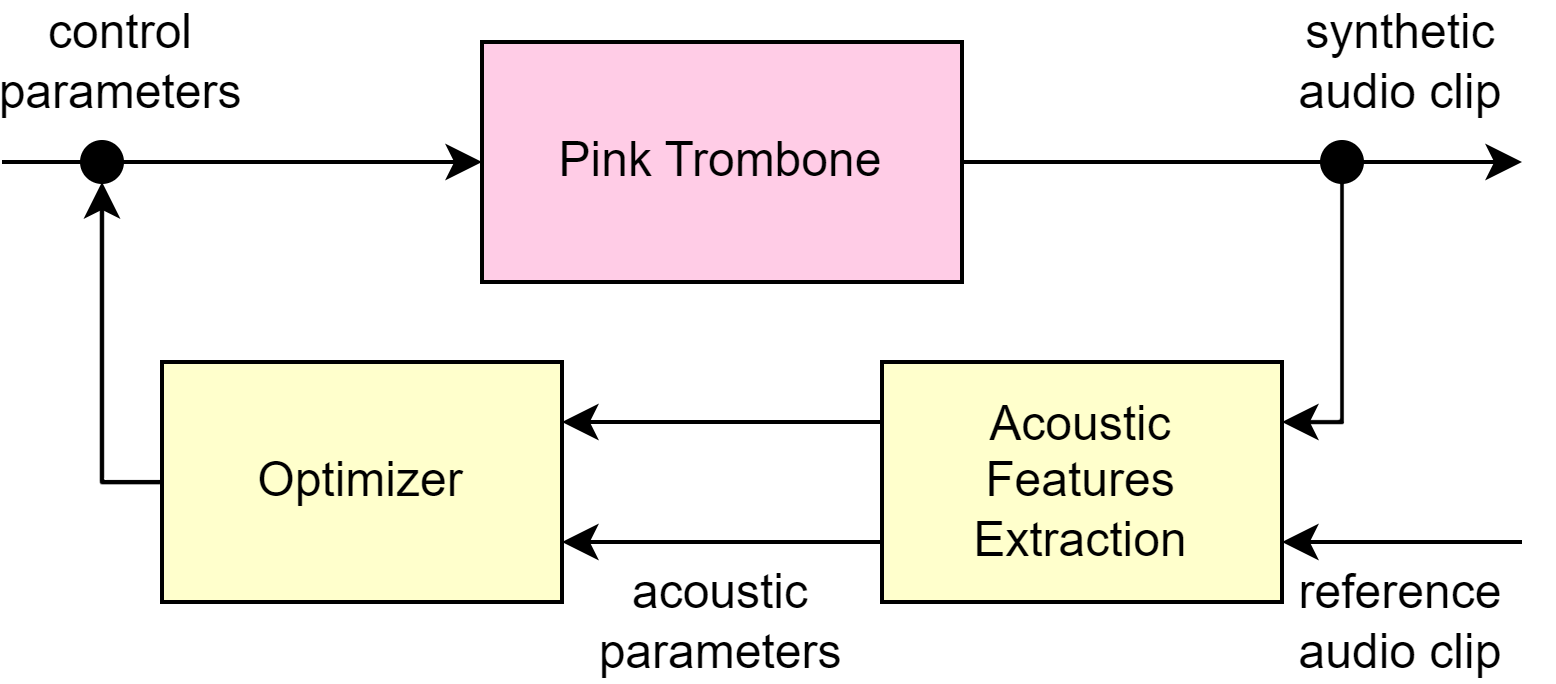}
\caption{\textit{Schematic on the optimization process.}}
\label{fig:PToptimizer}
\end{figure}

For sound-matching optimization, one may focus on the control parameters of the synthesizer, the input acoustic features extracted from the audio, and the process that leads to optimization. All these aspects provide insights into the methodologies and objectives of various optimization studies in synthesizers. Fig. \ref{fig:PToptimizer} depicts the overall schematic of the optimization process. 

\subsection{Optimization Methods}
Considering the complexity of sound synthesizers, there is a need for reliable optimization techniques. Numerous optimization methods have been investigated in terms of optimizing parameters for physical models or traditional synthesizers. The related work can be organized into two main categories:
\paragraph*{Search-based Methods:}{these have been widely applied to physical models in audio synthesis due to their ability to handle non-differentiable, non-linear, and non-convex optimization problems. They are universal and regard the synthesizer as a black-box model, focusing solely on parameter space optimization. Standard ways include the use of Evolutionary Algorithms (EA), including Evolution Strategies \cite{mitchell2007evolutionary}, Genetic Algorithm (GA) \cite{lai2006automated}, or Particle Swarm Optimization (PSO) \cite{zuniga2019realistic}. Other methods, including Hill Climber \cite{yee2018automatic}, Levenberg–Marquardt Algorithm \cite{pyz2012lithuanian} and Nelder-Mead Method \cite{baghmisheh2008frequency} are also considered.}

\paragraph*{Model-based Methods:}{machine learning (ML) models have become mainstream for synthesizer parameter estimation in recent years. They learn the mapping between the latter and audio features directly from data. In \cite{barkan2019deep}, authors used a strided Convolutional Neural Network (CNN) to predict the parameters of a subtractive synthesizer. Recent work proposed differentiable digital signal processing (DDSP) \cite{engel2020ddsp} and integrated an additive synthesizer with a filtered noise synthesizer into the end-to-end deep learning framework. These allow direct gradient descent optimization. DDSP is now widely utilized for parameter estimation \cite{masuda2021synthesizer}, despite its need for precise reproduction of the target synthesizer in a differentiable manner, which poses difficulties.}

Each approach has its own benefits and limitations, leading to ongoing discussions. In \cite{yee2018automatic}, authors compared sound-matching performance on a VST synthesizer using two search strategies and three neural network methods. 
Results indicated that search methods are limited by their computational cost, and modeling methods are restricted to the inductive bias of model structure and data availability. We tested these limitations for the PT, including simple speech and non-speech vocalisations to evaluate the performance of the optimized model parameters to reproduce sounds. 

\subsection{Parameter Selection}
The control parameters of the synthesizer and the input parameters for the optimizer largely depend on the synthesis technique and the desired outcomes, in accordance with Fig. \ref{fig:PToptimizer}. We focus on the PT control parameters --see Table \ref{tab:pt-params}. Physical models may alternatively use local constrictions to describe the configuration required for the vocal tract to produce a certain sound. The PT can actually operate on those as well. Nonetheless, we are interested in the primary ones. 

Furthermore, inputs to the optimizer must represent the acoustic content of the audio \textcolor{black}{samples} so that the model may produce accurate outputs. For this task, we shall look at acoustic features. 

\subsection{Acoustic Features Extraction}
Various acoustic features have been used in synthesizer optimization studies to evaluate and quantify the quality of the synthesized sounds to guide the optimization process. Spectral features are the most common, but finding the best metric with good perceptual consistency is still an open question \cite{moffat2018perceptual}. In \cite{munoz2011opposition} authors focus on the spectral norm error, \cite{lai2006automated} used spectral norm plus spectral centroid error extracted from short-time Fourier transform (STFT) for each frame, \cite{mitchell2007evolutionary} used relative spectral error, which is computed by summing normalized differences between frequency components extracted from two spectra, \cite{riionheimo2003parameter} combined the least squared error of the STFT of two sounds plus the perceptual error applying a narrow band masking curve. On error computation, \cite{yee2018automatic} used Euclidean distance of Mel-Frequency Cepstral Coefficients (MFCCs),  \cite{engel2020ddsp} used a deep representation extracted from MFCCs, and multiscale spectral loss plus perceptual loss, \cite{barkan2019deep} compared the following features as the Deep Neural Networks (DNN) input: a set of spectral features \cite{itoyama2014parameter}, STFT, and deep representation extracted by a CNN from the raw signal. Results showed that STFT and deep representations seem more representative than handcrafted features. Our aim now is to identify suitable ones for the PT parameters optimization.
 %

\section{EXPERIMENTATION}
We designed our experiments to focus on three specific characteristics that are relevant to the acoustic-to-articulatory inversion: the \textit{optimizer} used to predict control variables, the \textit{audio representation} to compute the difference between original and synthesized audio, and the \textit{signal complexity}. 

\subsection{Pink Trombone Fundamentals}

The PT is a simple vocal tract model that can be interacted with through a web interface. It is a Kelly-Lochbaum (KL) type model whose technical details can be found in \cite{kelly1962speech}. In our black-box case study, we focused on the number of parameters to be used and their bounds. Table \ref{tab:pt-params} summarizes these, which correspond to physical attributes of the vocal tract. We aimed to decouple the meaning of these parameters from their human meaningfulness for our method to be useful to any synthesizer.

\begin{table}[!h]
\caption{\emph{Pink Trombone parameters and their bounds.}\label{tab:pt-params}}
\begin{tabular}{lcc}
Pink Trombone Parameters   & Lower bound & Upper bound \\ \hline
Pitch (Hz)                     & 75          & 330         \\
Voiceness                  & 0           & 1           \\
Tongue Index               & 14          & 27          \\
Tongue Diameter (cm)       & 1.55        & 3           \\
Lips Diameter (cm)         & 0.6         & 1.2         \\
Constriction index         & 12          & 42          \\
Constriction Diameter (cm) & 0.6         & 1.2         \\
Throat Constriction (cm)   & 0.5         & 1.0        
\end{tabular}
\end{table}

\subsection{Signal complexity}
\label{subsec:signal-complexity}

Signal complexity refers to the challenges we pose to the optimizers to predict the exact parameters. In that sense, we consider three independent characteristics of the signal. First, the \textit{origin of the audio file}: audio generated by a speech synthesizer or a person. Second, \textit{variations over time}: sustained notes or dynamic audio (such as a yawn). Third, \textit{number of variables to optimize}: related to the characteristics of the synthesizer. Hereafter we enumerate all experiments conducted from the least to the most complex.


\begin{itemize}
    \item PT generated sounds for which:
    \begin{itemize}
        \item One of the control parameters is unknown.
        \item All of the control parameters are unknown.
        \item Gaussian white noise is added. This evaluates the robustness of optimizers dealing with non-ideal signals.
        \item Control parameters vary over time.
    \end{itemize}
        
    \item Audio \textcolor{black}{clips} containing: 
    \begin{itemize}
        \item Sustained vowel sounds.
        \item Yawnings.
    \end{itemize}
\end{itemize}

\subsection{Audio Representation and Quality Assessment}
\label{subsec:error-funcs}


\subsubsection{Representations focusing on spectral difference}
\label{subsubsec:spec-diff}
To minimize the difference between the target and reconstructed sound, we focused on the spectral features of the audio signals. The following list includes all the transformations evaluated:

\begin{itemize}
    \item \textit{STFT}. A window size of 1024 samples with a 2x overlap STFT was taken.
    \item \textit{Multiscale spectrogram}. The window sizes were \{64, 128, 256, 512, 1024\}, with a 75\% overlap. 
    \item \textit{MEL-spectrogram}. We used 128 filters in the MEL bank up to a maximum frequency of 8\,KHz.
    \item \textit{MFCCs}. We took 20 cepstral coefficients from the MEL-spectrograms.
\end{itemize}

Computations were performed in Python 3.9, using the AuraLoss library \cite{steinmetz2020auraloss}. The Mean Absolute Error (MAE) between the input  and reconstructed audio was computed as the error function.

\subsubsection{Perceptual metrics}
In addition to MAE, we also computed a set of perceptual quality and intelligibility metrics. These metrics were not used as error functions in the optimization process. The findings may be representative of the perceptual similarity between sounds. However, we encourage readers to listen to the results we posted online. The following full reference metrics were analyzed:
\begin{itemize}
    \item PESQ: Perceptual Evaluation of Speech Qlt. \cite{rix2001perceptual}. 
    \item PEAQ: Perceptual Evaluation of Audio Qlt. \cite{thiede2000peaq}. 
    \item ViSQOL: Visually-Inspired Speech Qlt. Obj. Listener \cite{chinen2020visqol}. 
    \item STOI: Short-Time Objective Intelligibility \cite{french1947factors}. 
\end{itemize}

\subsection{Selected Optimizers}
\label{subsec:optimizers}

We used \textit{optimization algorithms} and a \textit{CNN} to predict the control parameters of the synthesizer. We fed the algorithms with the MAE between the original and the synthesized signal. Hereafter we briefly introduce the selected optimization algorithms, which we have evaluated in terms of computational cost and reconstruction error.

\paragraph*{Genetic Algorithm (GA):} is an optimization technique inspired by natural selection and genetics \cite{holland1992genetic}. The candidate solutions are defined by a set of genes. In every generation (loop over all candidates), the genes are able to randomly change (mutation), combine with other candidates (crossover), and be selected (optimization) to search for optimal solutions in the solution space. The fitness function seeks to minimize differences in the input/output signals.


We used 32 bits to define the genes, a crossover rate of 0.9, a mutation rate of 0.03, and a population of 10 individuals.

\paragraph*{Particle Swarm Optimization (PSO):} is a nature-inspired metaheuristic optimization technique that simulates the social behavior of swarms  \cite{kennedy1995particle}. PSO operates by iteratively adjusting the position of particles within the search space based on their individual and global best experiences, converging towards the optimal solution. In our case, we set acceleration parameters $c_1=0.5$, $c_2=0.3$ (trust in itself, trust in its neighbors), and inertia weight $w=0.9$, with 10 particle population.
\textcolor{black}{
\paragraph*{Trust Region reFlective (TRF):}
The Trust Region reFlective \cite{branch1999subspace} algorithm is a computational technique for solving least squares optimization problems. It employs a model-based method, seeking to minimize a function by iteratively creating simplified models of the objective function within certain trusted regions. The term ``reflective'' refers to the method's way of handling boundaries and constraints: if a proposed step hits a boundary, it is reflected in the feasible region.}





\paragraph*{Nelder-Mead Method (NM):} also known as the downhill simplex method \cite{10.1093/comjnl/7.4.308}, is a multidimensional optimization technique well suited for non-linear problems. The algorithm starts with an initial simplex, a set of $n+1$ points in an $n$-dimensional space. The algorithm iteratively updates the position of the simplex by reflecting, expanding, contracting, or shrinking it, based on the values of the function being optimized at the vertices of the simplex.

\paragraph*{Covariance Matrix Adaptation Evolution Strategy (CMA-ES):} is a stochastic optimization algorithm that uses information about the distribution of the samples generated by the algorithm to guide the search for the optimal solution \cite{inproceedingsHansenAdpting1996}. The algorithm starts with an initial guess for the solution and then generates a set of samples around this point. The distribution of these samples is then updated based on the fitness of the samples, with higher-fitness samples being more likely to be selected. As the algorithm progresses, outcomes increasingly concentrate around the optimal solution.

\paragraph*{Trust Region Reflective (TRF):} is a nonlinear optimization algorithm \cite{nocedal1980updating}. The algorithm works by iteratively approximating the objective function with a quadratic model within a trust region around the current solution, then solving the subproblem within the trust region to find the next iterate. The trust region radius is adjusted based on the model's performance at each iteration.

\paragraph*{Neural network prediction (NN):} 
\label{subsec:nn}
In addition to the optimization techniques, we tested the capability of neural networks to predict the control variables based on the acoustic features. We designed a CNN that admits spectrograms as input and outputs the control variables. To train it we collected a database of 400,000 different PT \textcolor{black}{clips}. We trained four different networks, each admitting as input for each audio representation mentioned in subsection \ref{subsubsec:spec-diff}. The network has been coded with Pytorch 1.7.1, with 2 convolutional layers, ReLU as the activation function, 0.0001 as the learning rate, ADAM optimizer, and following the 60/20/20 data splitting strategy between training, validation, and test.

\subsection{Materials}
Attending to the scope of this contribution, the assessment of the performance achieved by the different techniques, audio representations, and optimizers in predicting the parameters of the physical model for sound-matching required two sets of audio files:
\begin{itemize}
    \item Synthetic \textcolor{black}{audio samples}, generated at 48\,kHz sampling rate and {1\,s} long. To generate these, we used the Programmable version of the PT\footnote{https://github.com/zakaton/Pink-Trombone} modified to be a Node.js server. \textcolor{black}{We generated 80 audio clips with random control parameters.}
    \item Audio \textcolor{black}{samples} downloaded from Freesound containing utterances from different speakers. We focused on sustained vowels (5 \textcolor{black}{clips}) and yawnings (8 \textcolor{black}{clips}). The vowels are one second long and the yawnings are three seconds on average. All files were recorded at a 48\,kHz sampling rate to match the same conditions as the synthetic \textcolor{black}{audio}. 
\end{itemize}

\section{RESULTS}

In this section, we present the results of the experiments aimed at predicting control parameters for PT. We sought to fix the same conditions for all optimizers to ensure a fair comparison. Some considerations apply to all experiments:

\begin{itemize}
    \item \textit{Error values are normalized} with respect to the maximum and minimum values that each parameter can take. 
    \item The \textit{random seed was fixed} to randomize the PT control parameters in each experiment, such that the optimizers face the same initial conditions in all cases.
    \item Each experiment was \textit{repeated 20 times}. Initial conditions and target values were randomized. 
    \item All optimizers had the same \textit{stop criterion}: reach an error of less than 0.0001 in the metric or stop to improve the relative error with respect to the previous 20 loops.
\end{itemize}

\subsection{Optimization of PT-generated sounds}
Hereafter we present the results of the different tests that were conducted using PT synthetic audio clips as inputs. 

\subsubsection{Optimization of one control parameter}

\begin{figure}[!t]
\centering
\includegraphics[width=68mm]{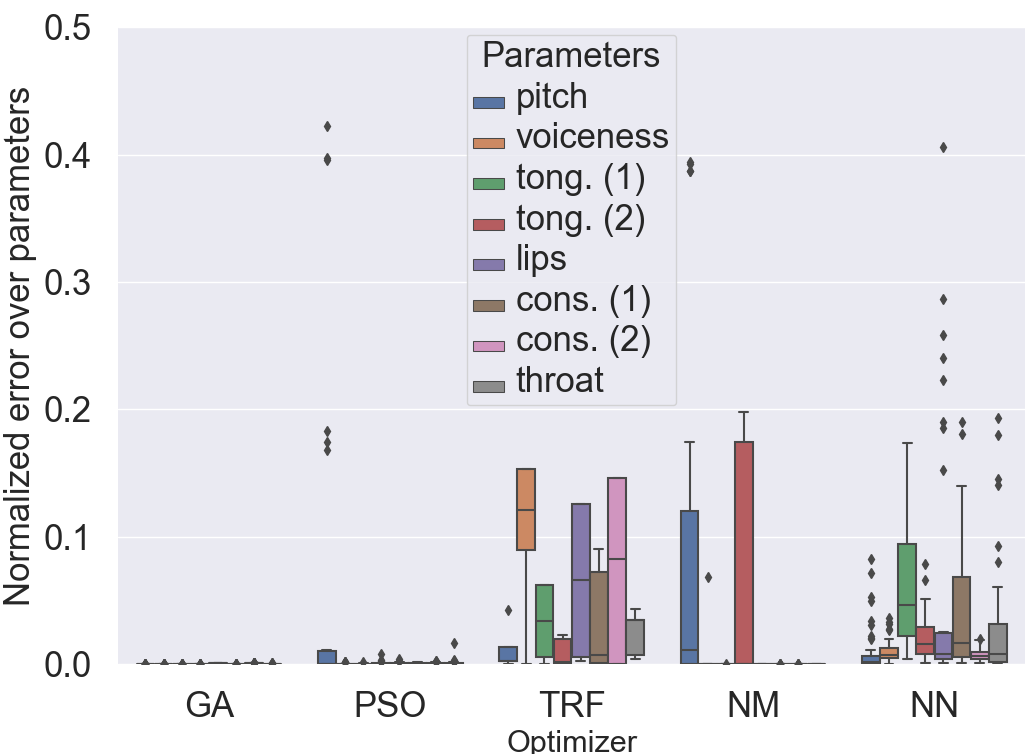}
\caption{\it Optimizer performance over one control parameter. X-axis includes the optimizers. Y-axis represents the normalized error. Each bar is a control parameter.}
\label{fig:independent_params_x_optimizers_y_error_hue_param}
\end{figure}

In this experiment, we fixed all control parameters except for one. We predicted the unknown value. 
This set of experiments does not include the CMA-ES algorithm because its particular design does not support single-parameter prediction. The results are shown in Figure \ref{fig:independent_params_x_optimizers_y_error_hue_param}, for the different optimizers and PT control parameters. 

Results demonstrate the effectiveness of GA and PSO in accurately predicting the control parameters. No outliers were observed in the genetic algorithm's performance. The NM algorithm successfully reached the absolute minimum for most parameters. However, it struggled to achieve the same for the pitch and one tongue-related parameter. A closer examination of these parameters revealed that their error functions contained multiple local minima. Since the performance of the NM is heavily influenced by its initial conditions, it makes it prone to getting stuck in them.

Despite not always arriving at the optimal values, TRF and NN converge rapidly to the minimum. Once it is trained, NN takes less than a second to reach the minima. TRF algorithm takes 5 seconds on average, which is four times faster to optimize than PSO and NM, and 100 times faster than GA.

\begin{figure}[!t]
\centering
\includegraphics[width=68mm]{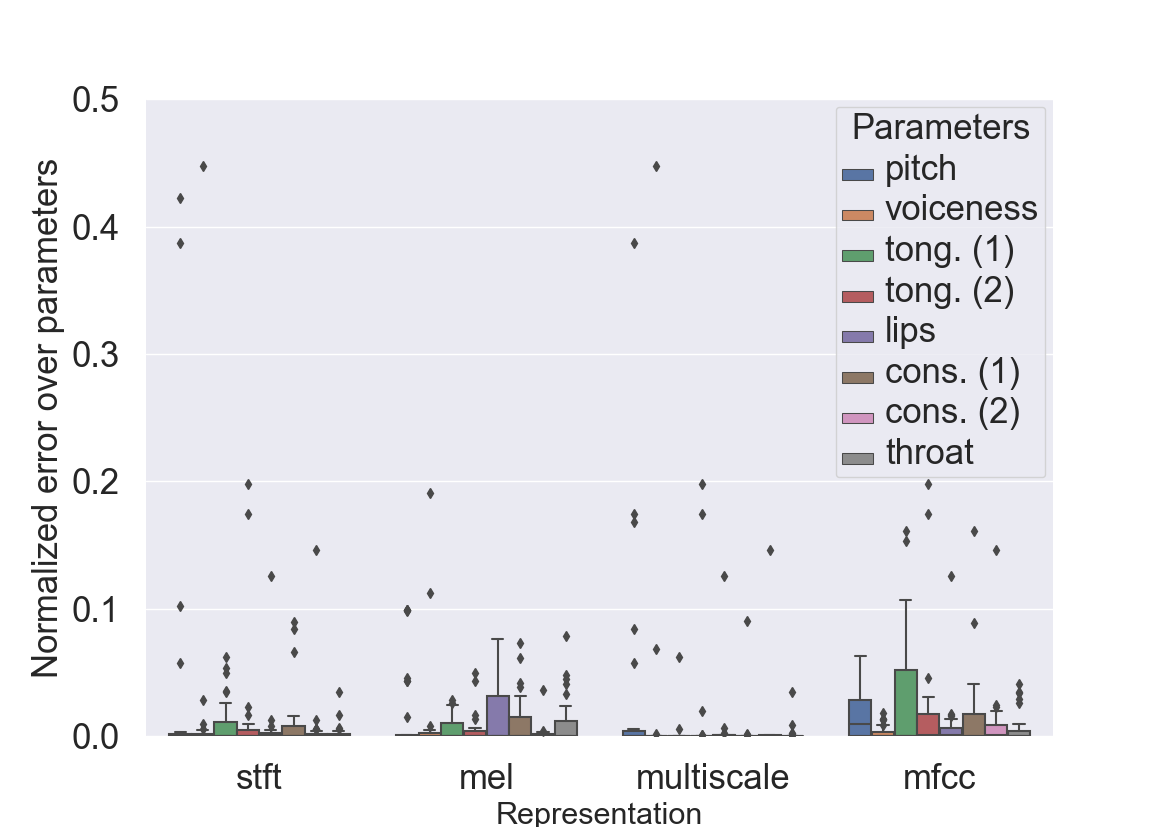}
\caption{\it Audio representation performance over one control parameter. X-axis represents each audio representation. Y-axis represents the normalized error. Each bar is a control parameter.}
\label{fig:independent_params_x_metric_y_error_hue_param}
\end{figure}

In the same line, Figure \ref{fig:independent_params_x_metric_y_error_hue_param} illustrates the error associated with each audio representation. All audio representations are suitable for optimizing individual control parameters. However, no error is observed in the multiresolution. This makes sense, since it is an extension of the STFT that better represents the spectral characteristics of the signal. 

\subsubsection{All control parameters}

\begin{figure}[!b]
\centering
\includegraphics[width=68mm]{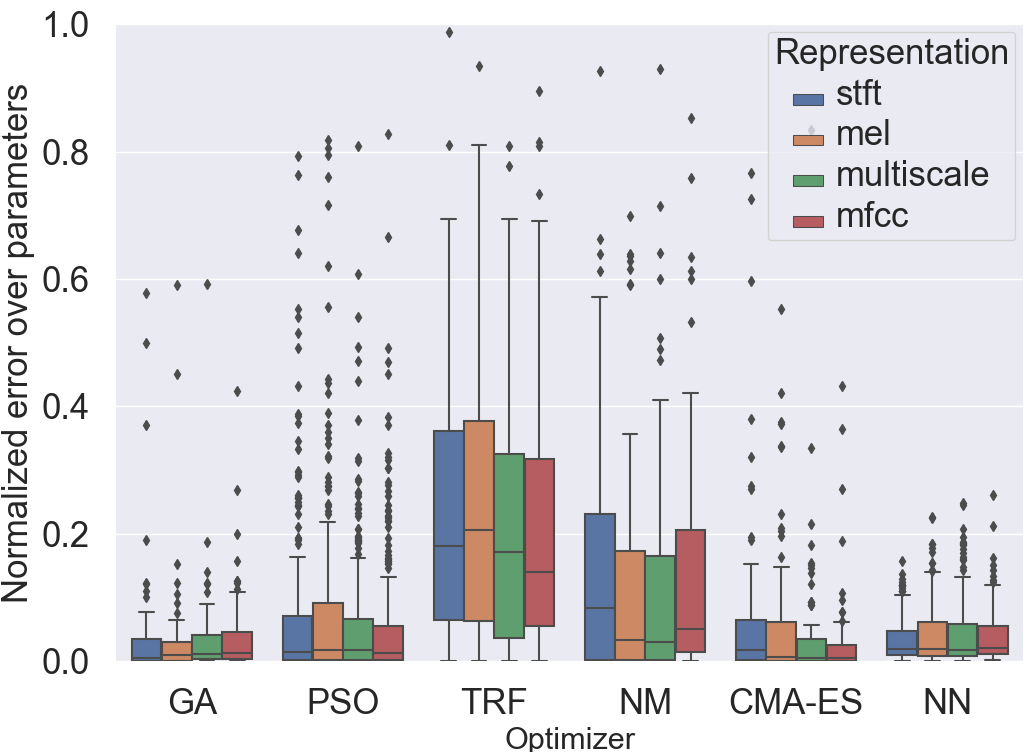}
\caption{\it Optimizer performance when all parameters are predicted at the same time. X-axis includes all optimizers. Y-axis represents the normalized error. Each bar corresponds to an audio representation.}
\label{fig:all_optimizers_x_param_y_error_hue_metric}
\end{figure}

The single-parameter experiments validate that articulatory parameters can be predicted from sound representations alone. However, this scenario is too simplified to clarify which optimizer is more accurate. This can be done by increasing the complexity of the experiment, seeking to predict all control parameters at once. The prediction results for each parameter are shown in Figure \ref{fig:all_optimizers_x_param_y_error_hue_metric}.

Experiments focusing on predicting all parameters demonstrate the superior performance of GA, CMA-ES, and PSO compared to other methods. In this experiment set, the eight-dimensional search area makes the optimization more challenging. The TRF and NM algorithms yielded unsatisfactory results, deeming them unsuitable for tackling the problem. As the number of potential solutions grows exponentially with increasing dimensions, only the most robust methods can find an optimal solution. Genetic algorithms and PSO can outperform least squares minimization or the downhill simplex method because they are more robust in handling complex search spaces, non-convex functions, and intricate relationships between variables.

\begin{figure}[!t]
\centering
\includegraphics[width=68mm]{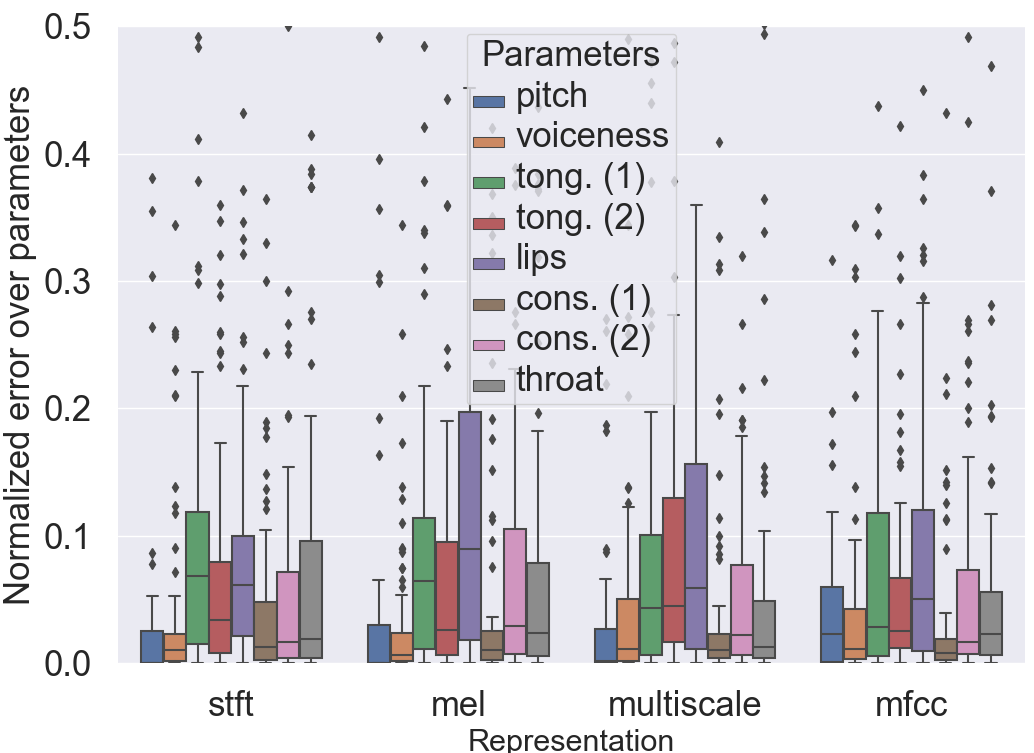}
\caption{\it Audio representation performance while predicting all parameters at a time. X-axis covers the representations. Y-axis represents the normalized error. Each bar corresponds to a control parameter.}
\label{fig:all_params_x_param_y_error_hue_metric}
\end{figure}

On the other hand, observing how the different audio representations behave in this scenario is interesting. They are shown in Figure \ref{fig:all_params_x_param_y_error_hue_metric}. We can observe that no representation performs significantly better than the rest, not even the multiresolution representation. However, finding a higher error is not necessarily a serious problem when reconstructing the signal. Most control parameters have local minima very close to the global minimum. This means that different PT configurations can produce almost the same reconstruction. This does not apply to the pitch parameter, which is one of the critical parameters in quality and, as can be seen, the MFCCs do not make it easy to reach its minimum. 

\begin{figure}[!b]
\centering
\includegraphics[width=68mm]{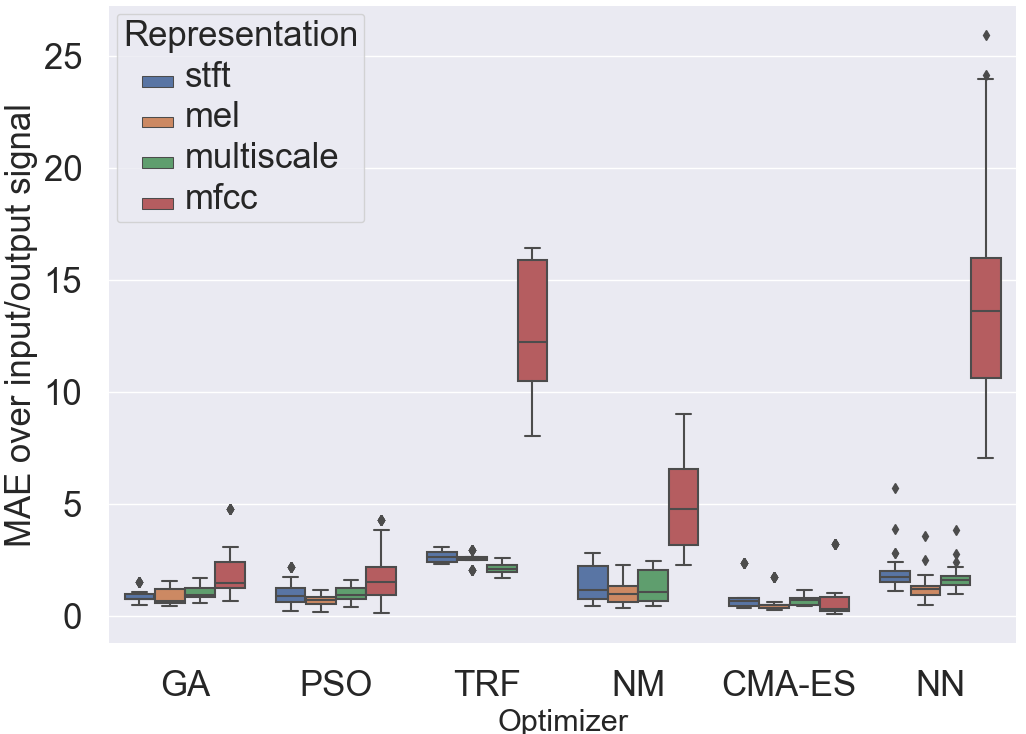}
\caption{\it Absolute performance of the optimizers and representations. X-axis includes all optimizers. Y-axis represents the MAE of the target and reconstructed audio file. Each bar is the audio representation.}
\label{fig:all_params_x_optimizer_y_globalerror_hue_metric}
\end{figure}

In fact, Figure \ref{fig:all_params_x_optimizer_y_globalerror_hue_metric} illustrates precisely these phenomena. It shows the MAE of the original and reconstructed signal. It is observed that regardless of the optimizer used when the search space is large, the MFCCs do not achieve satisfactory results. Thus, this experiment indicates that the best prediction of the control parameters can be made with GA or the PSO using the MEL scale or Multiresolution spectrograms. 

\begin{figure}[!t]
\centering
\includegraphics[width=68mm]{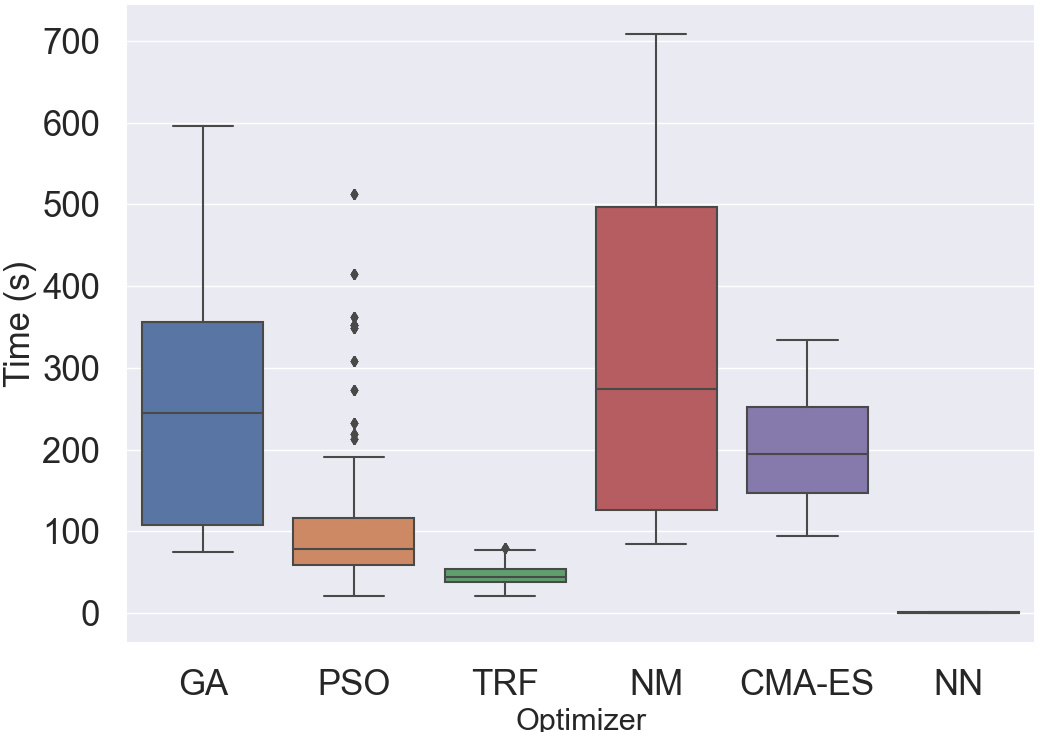}
\caption{\it Computational cost for the different optimizers --in X-axis. Y-axis represents the time to converge, in seconds.}
\label{fig:computational_cost}
\end{figure}

In addition, Figure \ref{fig:computational_cost} shows the computational costs of each optimizer. It shows that the NN is the fastest once trained, while PSO is the fastest of the suitable optimization techniques. 

\subsubsection{All control parameters which vary over time}

\begin{figure}[!t]
\centering
\includegraphics[width=68mm]{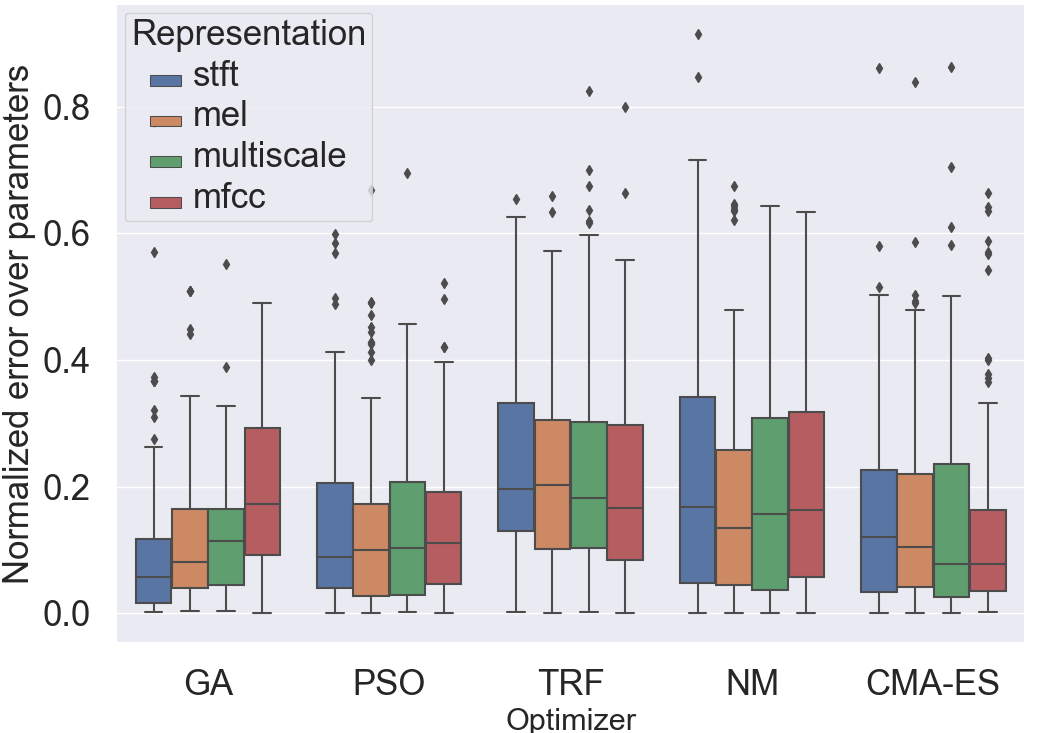}
\caption{\it Performance of optimizers and representations when the signal varies over time. X-axis includes all audio optimizers. Y-axis represents the normalized error between the target and reconstructed audio. Each bar corresponds to an audio representation.}
\label{fig:two_params}
\end{figure}

The next level of complexity we tested was optimizing parameters that varied over time. To achieve this, we created an interpolator that generated intermediate values between two temporal spaces defined by two sets of articulatory parameters in the PT. As shown in Figure \ref{fig:two_params}, none of the optimizers were able to achieve a satisfactory result when optimizing a time-variant set of parameters. The only optimizer that achieved a result closer to zero was the GA, using the STFT. The tendency is that as more parameters are added to optimize, the search space becomes more complex and therefore very difficult to reach the absolute minimum. It is important to note that this method is not suitable for a neural network. It would be necessary to train new networks depending on the number of parameters to be predicted. 

To achieve a more satisfactory, general solution, it was decided that the best strategy for optimizing signals that vary over time is to segment the signal into small windows and optimize them as if they were a static signals. We tested different window sizes and found out a 100 milliseconds length performed optimally. These windows can then be connected using a Savitzky-Golay filter \cite{doi:10.1021/ac60214a047}, which smooths out the result. The optimization results of these tests suggest insights equivalent to predicting a non-variant set of parameters. That is why this technique is the preferred choice for optimizing sounds created by humans.

\subsubsection{All control parameters in noise}

\begin{figure}[!b]
\centering
\includegraphics[width=68mm]{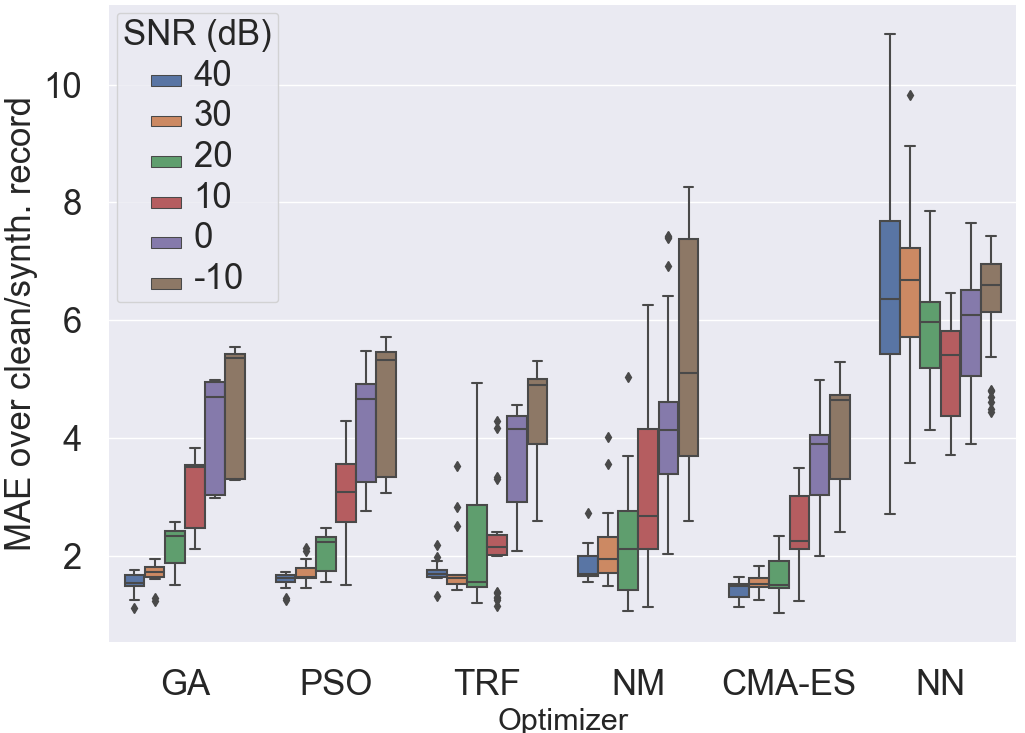}
\caption{\it Performance of the optimizers when Gaussian White Noise was applied at the input. X-axis includes allaudio optimizers. Y-axis represents the MAE between the target and the reconstructed audio file. Each bar is a different Signal-to-Noise ration.}
\label{fig:noise_plot}
\end{figure}

In these experiments, different amounts of Gaussian white noise were added to the original signal. We sought to predict the articulatory parameters that defined the signal. As shown in Figure \ref{fig:noise_plot}, the optimizers performance deteriorates as more noise is added. Additionally, we observed that the optimizers do not start to exhibit exponentially growing errors until the signal-to-noise ratio reaches 20 decibels. All optimizers act similarly, with the exception of the NN, which does not tolerate noise at its input.

From these experiments, we can conclude that it is possible to optimize signals that are not perfectly generated by a synthesizer but may come from any source, such as a recording from a public database. This finding is significant because it suggests that our approach can be applied in real-world scenarios where the input signals will likely not be perfectly recorded. 

\begin{table*}[!h]
\caption{{\it Perceptual equivalent metrics of the real sounds. Type "PT" stands for "Pink Trombone" generated, "VW" for sustained "Vowel", and "Y" for "Yawn". All metrics are in MOS scale (from 1 to 5) except STOI (from 0 to 1). We used a color code to indicate the best-performing set of acoustic parameters per audio type and optimizer. }}
\resizebox{\textwidth}{!}{%
\begin{tabular}{cccccccclccll}
\multicolumn{1}{l}{}     &    & GA                                    & PSO                                   & TRF                                   & NM                                    & CMA-ES                                & NN                                    & \multicolumn{1}{c}{} & Best   & Result                                &  & \multicolumn{1}{c}{Legend}                             \\ \cline{2-11} \cline{13-13}
                         & PT & \cellcolor[HTML]{D0CECE}$2.2 \pm 0.9$ & \cellcolor[HTML]{D9E1F2}$2.1 \pm 1.2$ & \cellcolor[HTML]{D0CECE}$1.8 \pm 1.0$   & \cellcolor[HTML]{E2EFDA}$2.2 \pm 1.0$   & \cellcolor[HTML]{D9E1F2}$2.6 \pm 0.9$ & \cellcolor[HTML]{D0CECE}$1.8 \pm 0.9$ &                      & CMA-ES & \cellcolor[HTML]{D9E1F2}$2.6 \pm 0.9$ &  & \multicolumn{1}{c}{\cellcolor[HTML]{D9E1F2}mel}        \\
                         & VW & \cellcolor[HTML]{D0CECE}$1.8 \pm 0.8$ & \cellcolor[HTML]{E2EFDA}$1.5 \pm 0.4$ & \cellcolor[HTML]{D9E1F2}$1.6 \pm 0.6$ & \cellcolor[HTML]{D9E1F2}$1.8 \pm 0.7$ & \cellcolor[HTML]{FCE4D6}$1.5 \pm 0.5$ & \cellcolor[HTML]{D0CECE}$1.8 \pm 1.2$ &                      & GA     & \cellcolor[HTML]{D0CECE}$1.8 \pm 0.8$ &  & \multicolumn{1}{c}{\cellcolor[HTML]{FCE4D6}mfcc}       \\
\multirow{-3}{*}{PESQ}   & Y  & \cellcolor[HTML]{E2EFDA}$1.5 \pm 0.4$ & \cellcolor[HTML]{D0CECE}$1.3 \pm 0.3$ & \cellcolor[HTML]{D9E1F2}$1.4 \pm 0.2$ & \cellcolor[HTML]{D9E1F2}$1.3 \pm 0.1$ & \cellcolor[HTML]{FCE4D6}$1.3 \pm 0.2$ & \cellcolor[HTML]{D0CECE}$1.5 \pm 0.7$ &                      & GA     & \cellcolor[HTML]{E2EFDA}$1.5 \pm 0.4$ &  & \multicolumn{1}{c}{\cellcolor[HTML]{E2EFDA}multiscale} \\ \cline{2-11}
                         & PT & \cellcolor[HTML]{D0CECE}$3.0 \pm 0.6$   & \cellcolor[HTML]{D9E1F2}$3.2 \pm 0.8$ & \cellcolor[HTML]{FCE4D6}$2.6 \pm 0.6$ & \cellcolor[HTML]{E2EFDA}$3.0 \pm 0.9$   & \cellcolor[HTML]{D9E1F2}$3.5 \pm 0.7$ & \cellcolor[HTML]{D9E1F2}$3.0 \pm 0.8$   &                      & CMA-ES & \cellcolor[HTML]{D9E1F2}$3.5 \pm 0.7$ &  & \multicolumn{1}{c}{\cellcolor[HTML]{D0CECE}stft}       \\
                         & VW & \cellcolor[HTML]{D9E1F2}$2.8 \pm 0.7$ & \cellcolor[HTML]{D0CECE}$2.2 \pm 0.1$ & \cellcolor[HTML]{FCE4D6}$2.6 \pm 0.7$ & \cellcolor[HTML]{FCE4D6}$2.5 \pm 0.7$ & \cellcolor[HTML]{FCE4D6}$2.5 \pm 0.6$ & \cellcolor[HTML]{E2EFDA}$2.6 \pm 0.7$ &                      & GA     & \cellcolor[HTML]{D9E1F2}$2.8 \pm 0.7$ &  &                                                        \\
\multirow{-3}{*}{PEAQ}   & Y  & \cellcolor[HTML]{FCE4D6}$2.5 \pm 0.8$ & \cellcolor[HTML]{E2EFDA}$3.0 \pm 1.1$   & \cellcolor[HTML]{FCE4D6}$2.7 \pm 0.7$ & \cellcolor[HTML]{E2EFDA}$2.6 \pm 0.7$ & \cellcolor[HTML]{D0CECE}$2.7 \pm 1.0$   & \cellcolor[HTML]{D9E1F2}$2.8 \pm 1.0$   &                      & PSO    & \cellcolor[HTML]{E2EFDA}$3.0 \pm 1.1$   &  &                                                        \\ \cline{2-11}
                         & PT & \cellcolor[HTML]{D0CECE}$3.1 \pm 0.9$ & \cellcolor[HTML]{D9E1F2}$3.4 \pm 0.8$ & \cellcolor[HTML]{D0CECE}$1.7 \pm 0.5$ & \cellcolor[HTML]{D9E1F2}$3.3 \pm 1.2$ & \cellcolor[HTML]{E2EFDA}$4.3 \pm 0.7$ & \cellcolor[HTML]{D9E1F2}$3.0 \pm 0.6$   &                      & CMA-ES & \cellcolor[HTML]{E2EFDA}$4.3 \pm 0.7$ &  &                                                        \\
                         & VW & \cellcolor[HTML]{FCE4D6}$1.9 \pm 0.5$ & \cellcolor[HTML]{D0CECE}$2.0 \pm 0.2$   & \cellcolor[HTML]{D0CECE}$2 \pm 0.3$   & \cellcolor[HTML]{E2EFDA}$1.9 \pm 0.3$ & \cellcolor[HTML]{D0CECE}$2.1 \pm 0.4$ & \cellcolor[HTML]{D0CECE}$1.8 \pm 0.4$ &                      & CMA-ES & \cellcolor[HTML]{D0CECE}$2.1 \pm 0.4$ &  &                                                        \\
\multirow{-3}{*}{ViSQOL} & Y  & \cellcolor[HTML]{FCE4D6}$2.1 \pm 0.1$ & \cellcolor[HTML]{D9E1F2}$2.1 \pm 0.1$ & \cellcolor[HTML]{E2EFDA}$2.3 \pm 0.3$ & \cellcolor[HTML]{E2EFDA}$2.3 \pm 0.4$ & \cellcolor[HTML]{D0CECE}$2.1 \pm 0.1$ & \cellcolor[HTML]{D9E1F2}$1.9 \pm 0.2$ &                      & TRF    & \cellcolor[HTML]{E2EFDA}$2.3 \pm 0.3$ &  &                                                        \\ \cline{2-11}
                         & PT & \cellcolor[HTML]{E2EFDA}$0.3 \pm 0.2$ & \cellcolor[HTML]{D9E1F2}$0.4 \pm 0.3$ & \cellcolor[HTML]{E2EFDA}$0.1 \pm 0.1$ & \cellcolor[HTML]{E2EFDA}$0.5 \pm 0.4$ & \cellcolor[HTML]{E2EFDA}$0.5 \pm 0.3$ & \cellcolor[HTML]{D9E1F2}$0.2 \pm 0.1$ &                      & CMA-ES & \cellcolor[HTML]{E2EFDA}$0.5 \pm 0.3$ &  &                                                        \\
                         & VW & \cellcolor[HTML]{E2EFDA}$0.1 \pm 0.1$ & \cellcolor[HTML]{E2EFDA}$0.1 \pm 0.0$   & \cellcolor[HTML]{D0CECE}$0.1 \pm 0.0$   & \cellcolor[HTML]{D9E1F2}$0.1 \pm 0.1$ & \cellcolor[HTML]{E2EFDA}$0.1 \pm 0.0$   & \cellcolor[HTML]{E2EFDA}$0.1 \pm 0.0$   &                      & CMA-ES & \cellcolor[HTML]{E2EFDA}$0.1 \pm 0.0$   &  &                                                        \\
\multirow{-3}{*}{STOI}   & Y  & \cellcolor[HTML]{FCE4D6}$0.3 \pm 0.1$ & \cellcolor[HTML]{FCE4D6}$0.3 \pm 0.1$ & \cellcolor[HTML]{FCE4D6}$0.3 \pm 0.1$ & \cellcolor[HTML]{E2EFDA}$0.3 \pm 0.1$ & \cellcolor[HTML]{D9E1F2}$0.3 \pm 0.1$ & \cellcolor[HTML]{D0CECE}$0.1 \pm 0.1$ &                      & -      & $0.3 \pm 0.1$                         &  &                                                       
\\ \cline{2-11}
\end{tabular}
\label{tab:best-audios}
}
\end{table*}

\subsection{Optimization of real audio files}

Results of the tests with real sounds can be seen in Table \ref{tab:best-audios}. The results for each perceptual metric are shown for the best-performing combination of the optimizer-representation pair. The columns detail the optimizers and the color shows the best audio representation for each case. We used different perceptual metrics to measure how similar the sounds generated by the synthesizer were to human-generated ones. We also include how the perceptual metrics behaved when predicting PT \textcolor{black}{samples}. These set up a benchmark to compare the upper limit that could be reached. Still, we encourage readers to visit our website, where we have published these audio files, and evaluate the quality themselves.

As shown in the table, CMA-ES and GA achieved superior results compared to other optimizers in terms of perceptual similarity in most of the cases. It is important to note that the MOS (Mean Opinion Score) equivalent results can still be considered low compared to the scale, as the sounds are synthesized by a certain vocal tract that may not correspond to the vocal tracts of the people who generated the original sound. Therefore, we do not claim that we can produce an exactly identical sound but an equivalent one. 

For all types of signal, we used the strategy of dividing the signal into small windows and smoothing them out into full-length signals. Systematically, PT-generated sounds are predicted with better scores than human-generated sounds. Furthermore, in many of the cases we found that yawns are perceptually recognized as more similar than sustained vowels. This is because the timbre in the sustained vowel has a much greater influence than in the yawn. The PT has vocal characteristics that do not match those of the person who recorded the sounds. For this reason, it is more difficult to recreate a perfectly harmonic voice like the vowel than a noisy sound like the yawn. This does not imply that our optimizers are malfunctioning, as the goal is to create comparable sounds, not exactly the same. The STOI metric in this regard is very representative of this situation, giving the yawn almost equal score to the PT-generated values, while the vowel is perceived as different.

These experiments also yield two interesting insights. First, one may identify optimizer-representation combinations that perform better than others. In particular, multiscale representation works well for yawns, while for sustained vowels STFT representation can do the job. None performed well using MFCC. Thus, one may need to take into account the type of signal to get good results from the optimizer. Second, there is consistency among the perceptual metrics. Those experiments that are more challenging consistently score worse than simpler ones.

\section{CONCLUSION}
Optimization techniques effectively predict the parameters of the Pink Trombone to produce human-like vocalisations. The selected algorithms delivered tuned control parameters while operating on different acoustic features and metrics. The resulting \textcolor{black}{audio samples} match the selected input sounds regarding the absolute error and perpetual equivalent metrics. A similar trend was observed on sustained vowels and yawnings; as well as under additive noise conditions. Nonetheless, the lower performance levels for the collected audio \textcolor{black}{samples} compared to the synthetic inputs, in absolute error and according to the perceptual-equivalent metrics, may be influenced by the limited ability of the Pink Trombone to match sounds out of its standard tract setting.

We comprehensively evaluated some of the most commonly used optimization algorithms in a black-box approach, predicting their control parameters to synthesize non-speech sounds. We tested different audio representations and conducted experiments in different scenarios, ranging from simple single-parameter predictions to complex, time-varying parameters or non-speech human-made sounds. Our results show that the Evolution Strategies (GA and CMA-ES) \textcolor{black}{and Particle Swarm Optimization} with multiresolution representation are the most effective for predicting control parameters with minimum error (MAE $<1\%$) and high quality (ViSQOL 4.3 for PT, PEAQ 3.0 for yawns). \textcolor{black}{Also, PSO} achieved the best performance vs. computational cost ratio.

According to our results, GA and PSO algorithms were superior to the other optimization methods in most cases. The NM algorithm struggled with local minima, and the TRF algorithm, although fast, could not optimize the parameters satisfactorily. The NN could predict the control parameters when the input was a sound generated by the PT. However, it failed when confronted with real sounds or noisy inputs. NN strategy can be useful for certain scenarios since it is also very fast once trained, but it can hardly reach the generalization of the GA.

Regarding audio representations, our experiments demonstrate that all those that we tested, including MFCC, STFT, MEL, and multiresolution decomposition, are suitable for optimizing individual control parameters. However, the MFCC representation showed poorer pitch prediction capability than other representations. This is consistent with what is expected from a cepstral representation according to the literature. 

Perceptual metrics validate that the optimizers are able to faithfully predict \textcolor{black}{audio samples} generated by the synthesizer itself. Taking this benchmark, we can observe that real sounds do not reach such a high performance. The conclusion is that our synthesizer cannot achieve certain characteristics of real voices in the given conditions (e.g. vocal tract size). Nevertheless, comparable sounds have been achieved, which is the goal of our research.

Future research should explore new techniques that enhance the prediction of time-varying signals. Further analysis is needed to investigate the parameter's flow, whether it aligns with the typical structures of a human vocal tract or the chosen optimization strategy. Additionally, hierarchical optimizations can improve the neural network's performance. Predictions should be conducted in two stages by initially narrowing the bounds and then fine-tuning. 

Finally, future work should use this framework to benchmark other solutions to the problem, including alternative optimization methods, acoustic features, metrics, and subjective tests.

\section{Acknowledgments}

The authors would like to thank David Südholt for the valuable discussions, support and help. Activities described in this contribution were partially funded by the European Union's Horizon 2020 Research and Innovation Programme under grant agreement No. 101003750, the Ministry of Economy and Competitiveness of Spain under grant PID2021-128469OB-I00, and the UPM Research Programme, \textit{Programa Propio de I+D+I 2022}.

\bibliographystyle{IEEEbib}
\bibliography{pinktrombonepaper} 

\end{document}